# Towards Efficient and Multifaceted Computer-assisted Pronunciation Training Leveraging Hierarchical Selective State Space Model and Decoupled Cross-entropy Loss

**Fu-An Chao, Berlin Chen**

Department of Computer Science and Information Engineering

National Taiwan Normal University

{fuann, berlin}@ntnu.edu.tw

## Abstract

Prior efforts in building computer-assisted pronunciation training (CAPT) systems often treat automatic pronunciation assessment (APA) and mispronunciation detection and diagnosis (MDD) as separate fronts: the former aims to provide multiple pronunciation aspect scores across diverse linguistic levels, while the latter focuses instead on pinpointing the precise phonetic pronunciation errors made by non-native language learners. However, it is generally expected that a full-fledged CAPT system should perform both functionalities simultaneously and efficiently. In response to this surging demand, we in this work first propose **HMamba**, a novel CAPT approach that seamlessly integrates APA and MDD tasks in parallel. In addition, we introduce a novel loss function, decoupled cross-entropy loss (**deXent**), specifically tailored for MDD to facilitate better-supervised learning for detecting mispronounced phones, thereby enhancing overall performance. A comprehensive set of empirical results on the speechocean762 benchmark dataset demonstrates the effectiveness of our approach on APA. Notably, our proposed approach also yields a considerable improvement in MDD performance over a strong baseline, achieving an F1-score of 63.85%. Our codes are made available at https://github.com/Fuann/hmamba

Figure 1: A running example depicts the evaluation differences between APA and MDD systems in the reading-aloud scenario.

## 1 Introduction

In this era of globalization and technologization, computer-assisted pronunciation training (CAPT) systems have emerged as an appealing alternative to meet the pressing need for second language (L2) learning. In comparison with traditional curriculum learning, CAPT offers advantages in both time efficiency and cost-effectiveness. More critically, it shifts the conventional pedagogical paradigm from teacher-directed to self-directed learning, thereby providing a stress-free environment for L2 learners (Eskenazi et al., 2009). In addition, CAPT applications have achieved marked success in various commercial sectors and testing services, such as the APPs of Duolingo (McCarthy et al., 2021) and the SpeechRater (Zechner et al., 2009) developed by Educational Testing Service (ETS). Typically, a de-facto archetype system for CAPT encompasses a "reading-aloud" scenario, where a non-native speaker is given a text prompt and instructed to pronounce it correctly. In this context, previous literature broadly divides applications of CAPT into two categories: automatic pronunciation assessment (APA) and mispronunciation detection and diagnosis (MDD), with each category dedicated to specific facets of pronunciation training. APA aims to evaluate the spoken proficiency of L2 learners by providing fine-grained feedback on various aspect assessments (e.g., accuracy and fluency) across multiple linguistic levels (e.g., word and utterance level) (Kheir et al., 2023). To assess L2 learners' spoken proficiency, APA systems typically employ scoring models that are either jointly trained (Gong et al., 2022; Chao et al., 2022) or leverage multiple regressors in an ensemble paradigm (Bannò et al., 2022a; Bannò and Matassoni, 2022b) to generate scores across various aspects. As such, users can receive multi-aspect assessment scores predicted by an APA system, as illustrated in the example shown in Figure 1. In contrast to APA, MDD focuses more on non-native speakers' phonetic pronunciation errors (Chen and Li, 2016). These errors usually have clear-cut distinctions between correct and incorrect ones, and can be easily quantified through deletions, substitutions, and insertions. For instance, a number of MDD models are designed to capitalize on classifier-based modeling (Truong et al., 2004; Strik et al., 2009; Harrison et al. 2009), enabling precise identification of the exact positions where pronunciation errors occur within an utterance. This capability provides L2 learners with specific feedback on discrepancies between intended pronunciation and actual pronunciation.

Albeit the phonetic (segmental) errors are crucial in the beginning stages of non-native language learning, prosodic (suprasegmental) errors may often cause a detrimental impact on the perception of fluency and lead to poor intelligibility (Chen and Li, 2016). This effect may be more pronounced in learning stress-timed languages like English, especially for a learner whose mother tongue is a syllable-timed language, such as Chinese (Ding and Xu, 2016). To tackle this problem, APA can play a pivotal role by offering prosodic or intonation assessment for L2 learners. For example, Lin et al. (2021a) introduced rhythm rubrics to predict the traits of sentence-level stress in L2 English utterances, demonstrating a strong correlation with the prosody scores assessed by the human experts. In addition, Arias et al. (2010) proposed text-independent systems for assessing intonation and stress, focusing on measuring the similarity between a test-taker's intonation or stress curve and that of a reference response.

On these grounds, it is evident that both APA and MDD are indispensable ingredients of CAPT, playing complementary roles in its success. However, previous studies on APA and MDD appear to have developed independently, with limited research exploring their synergetic use. Ryu et al. (2023) proposed a joint model for APA and MDD, leveraging knowledge transfer and multi-task learning. Their findings indicate high negative correlations between several assessment scores and mispronunciations. This also suggests that the human assessors may be influenced by phonetic errors when evaluating overall proficiency scores for various aspects, which to some extent gives away the halo effect present in the human annotations. While the corresponding results show that jointly modeling both tasks can achieve better performance than modeling each task in isolation, only utterance-level holistic assessments are considered in their experiments. In order to provide more comprehensive and fine-grained feedback for L2 learners, other granularities, such as the phone level or the word level, should also be aptly modeled. Recognizing this importance, we propose HMamba, a more effective approach, for multifaceted CAPT. Being aware of the linguistic hierarchy, HMamba can capture the intrinsic multi-layered speech structure, delivering both coarse and fine-grained pronunciation assessments while offering more accurate diagnostic feedback of mispronunciations. In addition, to address the extra computational costs introduced by multi-task learning, HMamba leverages a selective state space model (SSM) that can efficiently tackle both APA and MDD tasks in parallel. The main contributions of this paper can be summarized as follows:

- We introduce HMamba, a unified and linguistically hierarchy-aware model that jointly tackles APA and MDD tasks, achieving superior overall performance compared to prior arts that employ either single-task or multi-task models.
- We propose a novel loss function, decoupled cross-entropy loss (termed deXent), which effectively addresses the inherent issue of text prompt-aware MDD methods. Notably, deXent is feasible and well-suited for optimizing the MDD performance, particularly in striking the balance between precision and recall.
- To the best of our knowledge, this is the first work to adopt and extend Mamba in the APA and MDD tasks for a more efficient and comprehensive CAPT application.

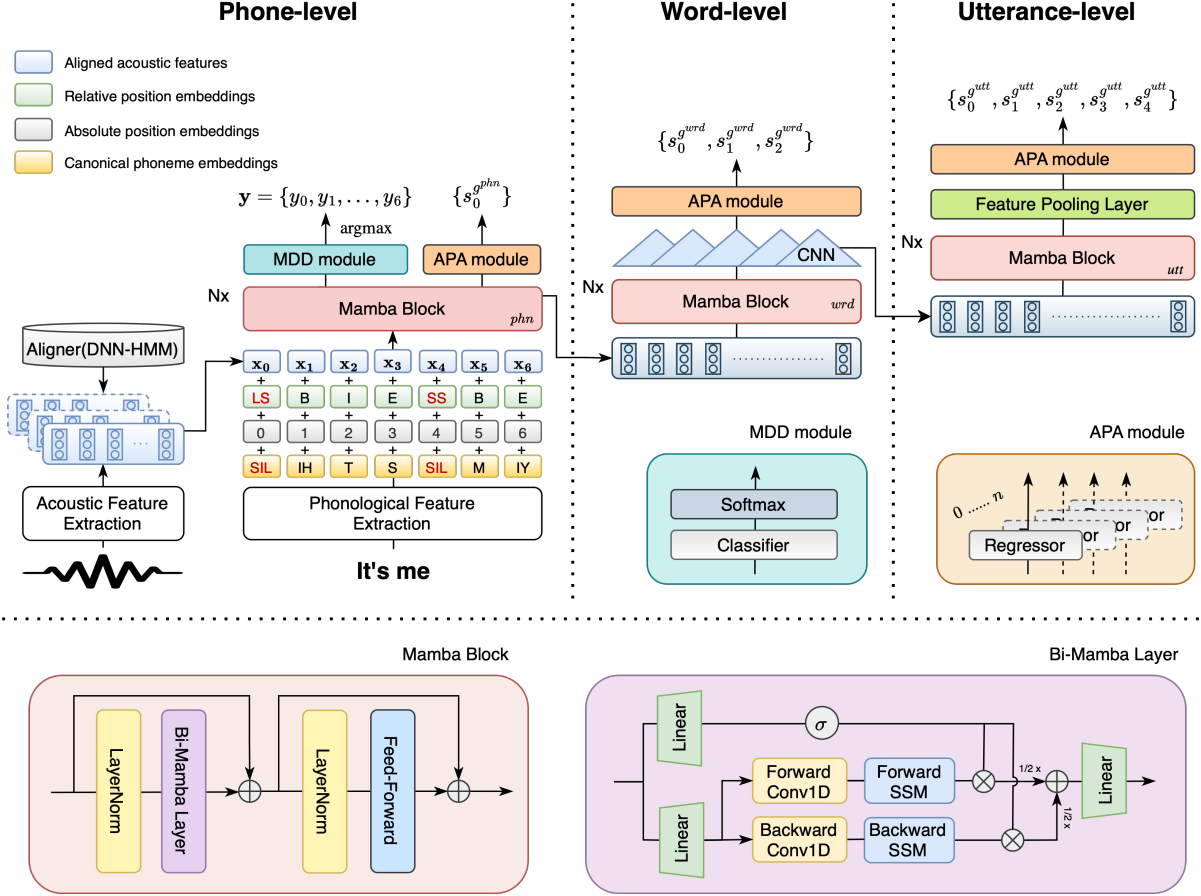

Figure 2: An overall architectural overview of HMamba, which consists of a bottom-up hierarchical modeling structure with several Mamba blocks across three levels (viz. phone, word, and utterance levels) that can perform multi-granular APA and MDD in parallel.

## 2 Methodology

### 2.1 Problem Definition

Considering an input time sequence of speech signal $\mathbf{u}$ uttered by an L2 learner and a reference text prompt $\mathbf{p}$ that contains $N$-length canonical phone sequence $\mathbf{p} = \{p_0, p_1, \ldots, p_{N-1}\}$, we adopt a set of feature extractors along with an aligner to extract an acoustic feature sequence $\mathbf{X} = \{\mathbf{x}_0, \mathbf{x}_1, \ldots, \mathbf{x}_{N-1}\}$ that aligned with $\mathbf{p}$ from $\mathbf{u}$. Our model aims to address APA and MDD tasks simultaneously but with separate processing flows: First, we define $G$ as a set of linguistic granularities, and for each granularity $g \in G$ the model manages to predict a set of aspect scores $\mathbf{s}^g = \{s_0^g, s_1^g, \ldots, s_{M_g-1}^g\}$, where $M_g$ refers to the number of aspect scores of target granularity $g$. In this work, $G = \{g^{phn}, g^{wrd}, g^{utt}\}$, where we have granularities of $g^{phn}$ (phone level), $g^{wrd}$ (word level), and $g^{utt}$ (utterance level) for the APA task. Meanwhile, the model also requires to detect error states $\mathbf{e} = \{e_0, e_1, \ldots, e_{N-1}\}$ with respect to $\mathbf{p}$ and in turn generate the correct diagnostic output $\mathbf{y} = \{y_0, y_1, \ldots, y_{N-1}\}$, where $y_n$ denotes the uttered phone of the learner corresponds to $p_n$.

### 2.2 HMamba

In this subsection, we shed light on the details of the proposed model, HMamba, which is devised as a hierarchical structure built upon the paradigm of selective SSM. A schematic illustration of the complete architecture is depicted in Figure 2. Specifically, HMamba synthesizes the APA and MDD modules, each of which contains multiple regressors and a classifier, respectively. These modules collectively generate the corresponding aspect score sequence $\mathbf{s}^g$ for each linguistic granularity $g$, as well as the phonetic error states $\mathbf{e}$ and diagnosis $\mathbf{y}$. Furthermore, each classifier and regressor is implemented with a simple feed-forward network (FFN) and jointly optimized through the training.

**Acoustic Feature Extraction:** In order to portray the non-native speaker's pronunciation quality, previous studies on either APA or MDD generally adopt a pre-trained acoustic model to extract goodness of pronunciation (GOP)-based features (Witt and Young, 2000; Hu et al., 2015; Shi et al., 2020). However, these features merely offer the segmental-level information that may not capture prosodic errors of an L2 learner. Given this limitation, we first utilize a pre-trained acoustic model as an aligner to identify phone boundaries (including silence), facilitating the extraction of other prosodic features such as the phone duration and statistics of root mean squared energy (Dong et al., 2024). To mitigate the low-resourced data problem (Chao et al., 2022), we also consider other self-supervised learning (SSL) features including wav2vec 2.0[1] (Baevski et al., 2020), HuBERT[2] (Hsu et al., 2021), and WavLM[3] (Chen et al., 2022). All these features are concatenated and subsequently projected through a linear layer to form a sequence of acoustic features $\mathbf{X}$. The transformation of each time step $t$ is given by

$$\mathbf{a}_t = [\mathbf{a}_t^{gop}; \mathbf{a}_t^{dur}; \mathbf{a}_t^{eng}; \mathbf{a}_t^{w2v}; \mathbf{a}_t^{hbt}; \mathbf{a}_t^{wlm}] \quad (1)$$

$$\mathbf{x}_t = \mathbf{W}\mathbf{a}_t + \mathbf{b} \quad (2)$$

where $\mathbf{W}$ and $\mathbf{b}$ are trainable parameters. Notably, a dropout rate of 10% is applied to all SSL features prior to the concatenation due to the discrepancy in dimensionality between these and other features.

**Phonological Feature Extraction:** In addition to acoustic cues, a common practice in CAPT is to inject the phonological information by introducing the reference text prompt features such as canonical phoneme embeddings (Gong et al., 2022), context-aware sup-phoneme embeddings (Chao et al., 2023), and vowel/consonant embeddings (Fu et al., 2021). In contrast to previous studies (Gong et al., 2022; Chao et al., 2022; Do et al., 2023a), we extract the canonical phoneme embeddings $\mathbf{E}^{phn}$ from $\mathbf{p}$ using a phone embedding layer that includes the silence (`SIL`) information which has been shown to be crucial when evaluating ones' spoken proficiency. Additionally, an absolute positional embedding $\mathbf{E}^{abs}$ and a relative position embedding $\mathbf{E}^{rel}$ are extracted. Distinct from $\mathbf{E}^{abs}$, $\mathbf{E}^{rel}$ denotes relative positions of phones in a word using tokens such as begin `[B]`, internal `[I]`, end `[E]`, and single-phone word `[S]` tokens. For special cases of silence positions, we explicitly categorize them as either long silence `[LS]` or short silence `[SS]`. Following the guideline suggested by ETS (Evanini et al., 2015), positions with a silence duration exceeding 0.495 seconds are assigned to `[LS]`; otherwise, they are assigned to `[SS]`. Finally, all these embedding features are point-wise added to $\mathbf{X}$ to obtain phone-level input features for subsequent modeling:

$$\mathbf{H}_0^{g_{phn}} = \mathbf{X} + \mathbf{E}^{phn} + \mathbf{E}^{abs} + \mathbf{E}^{rel} \quad (3)$$

The details of the complete feature ablations (both acoustic and phonological features) are shown in Appendix B.

**Mamba Blocks:** To foster highly efficient multi-task learning, we introduce selective SSMs instead of attention-based models such as the Transformer (Vaswani et al., 2017). Specifically, we adopt Mamba (Gu and Dao, 2023) as our backbone model structure in this work. Different from previous SSM instantiations, Mamba features an input-dependent selection mechanism and a hardware-aware algorithm, allowing for efficient input information filtering by dynamically adjusting the SSM parameters based on the input data. This also facilitate faster recurrent computation of the model using scan. Nevertheless, the vanilla Mamba conducts causal computations in a unidirectional manner, which prevents it from capturing global information as effectively as the multi-head self-attention (MHSA) module involved in Transformer. To address this problem, we explore a bidirectional variant of Mamba as the basic modeling block. In this approach, we replace the MHSA module in the Transformer encoder with a bidirectional Mamba layer, as depicted in Figure 2. Specifically, for input $\mathbf{H}^{g_i}$ to the Mamba block at granularity level $g$, the output $\mathbf{H}^{g_{i+1}}$ of the block is:

$$\mathbf{H}'^{g_i} = \text{BiMamba}(\text{LayerNorm}(\mathbf{H}^{g_i})) + \mathbf{H}^{g_i} \quad (4)$$

$$\mathbf{H}^{g_{i+1}} = \text{FFN}(\text{LayerNorm}(\mathbf{H}'^{g_i})) + \mathbf{H}'^{g_i} \quad (5)$$

---
[1] https://huggingface.co/facebook/wav2vec2-large-xlsr-53
[2] https://huggingface.co/facebook/hubert-large-ll60k
[3] https://huggingface.co/microsoft/wavlm-large

where BiMamba denotes the bidirectional Mamba layer and FFN refers to the feed-forward module, respectively. Notably, there are several studies investigating the bidirectional processing of Mamba (Liang et al., 2024; Zhang et al., 2024; Jiang et al., 2024). In this work, we use a similar structure as Jiang et al. (2024) to implement the bidirectional Mamba layer. For input $\mathbf{N}^{g_i}$ from the output of layer normalization of $\mathbf{H}^{g_i}$ to a bidirectional Mamba layer, the corresponding output $\mathbf{M}^{g_i}$ is computed as follows:

$$\mathbf{Z}^{g_i} = \text{Linear}(\mathbf{N}^{g_i}) \quad (6)$$

$$\mathbf{S}^{g_i \rightarrow} = \text{Linear}(\mathbf{N}^{g_i}), \quad \mathbf{S}^{g_i \leftarrow} = \text{Flip}(\mathbf{S}^{g_i \rightarrow}) \quad (7)$$

$$\begin{cases} \mathbf{C}^{g_i \rightarrow} = \text{Conv1D}^{\rightarrow}(\mathbf{S}^{g_i \rightarrow}) \\ \mathbf{C}^{g_i \leftarrow} = \text{Conv1D}^{\leftarrow}(\mathbf{S}^{g_i \leftarrow}) \end{cases} \quad (8)$$

$$\begin{cases} \mathbf{O}^{g_i \rightarrow} = \sigma(\mathbf{Z}^{g_i}) \bigotimes \text{SSM}^{\rightarrow}(\mathbf{C}^{g_i \rightarrow}) \\ \mathbf{O}^{g_i \leftarrow} = \sigma(\mathbf{Z}^{g_i}) \bigotimes \text{SSM}^{\leftarrow}(\mathbf{C}^{g_i \leftarrow}) \end{cases} \quad (9)$$

$$\mathbf{M}^{g_i} = \text{Linear}(\frac{1}{2}\mathbf{O}^{g_i \rightarrow} + \frac{1}{2}\text{Flip}(\mathbf{O}^{g_i \leftarrow})) \quad (10)$$

where $\mathbf{S}^{g_i \rightarrow}$ and $\mathbf{S}^{g_i \leftarrow}$ denote the forward and backward sequence features, respectively. Specifically, $\mathbf{S}^{g_i \leftarrow}$ is derived from $\mathbf{S}^{g_i \rightarrow}$ by a flipping operation $\text{Flip}(\cdot)$. $\text{Conv1D}(\cdot)$, $\sigma(\cdot)$, and $\text{SSM}(\cdot)$ represents the 1-D convolution, activation function, and selective SSM algorithm described in Mamba (Gu and Dao, 2023), respectively.

**Hierarchical Mamba:** Since the speech signals are typically distinguished by the complex hierarchical composition, prior studies (Do et al., 2023a; Chao et al., 2023) have suggested that hierarchical modeling structures is more amenable than parallel modeling structures (Gong et al., 2022). To capture the linguistic hierarchy while retaining the cross-aspect relations within the same linguistic unit, we design and instantiate our model with a hierarchical structure and introduce Mamba blocks to model the dependencies at each granularity level. More concretely, our approach generates finer granularity scores at the lower layers and coarser granularity scores at the higher layers, as exhibited in Figure 2. In phone-level modeling, we first use $\mathbf{H}^{g_0^{phn}}$ as the input into $L_p$-layer Mamba blocks to obtain the phone-level contextualized representations $\mathbf{H}^{g_{L_p}^{phn}}$:

$$\mathbf{H}^{g_{L_p}^{phn}} = \text{MambaBlock}_{phn}(\mathbf{H}^{g_0^{phn}}) \quad (11)$$

Subsequently, $\mathbf{H}^{g_{L_p}^{phn}}$ are then propagated forward into the APA module and the MDD module for solving a regression and a sequence classification problem, respectively. The APA module contains one regressor that aims to predict the phone-level aspect score $s_0^{g^{phn}}$ (accuracy). On the other hand, the MDD module comprises a classifier and a softmax function that cooperatively learn a distribution $\hat{y}_t$ over the phoneme classes $C$ for each time step $t$. The diagnosis $y_t$ can then be identified by applying the argmax function to $\hat{y}_t$. In this work, we streamline the MDD task by treating it as a process of free phone recognition (Li et al., 2015). As such, we can directly detect the corresponding error state $e_t$ by comparing $y_t$ with $p_t$, eliminating the need for a separate detection module. Meanwhile, the resulting $\mathbf{H}^{g_{L_p}^{phn}}$ is served as $\mathbf{H}^{g_0^{wrd}}$ for subsequent modeling.

In word-level modeling, $L_w$-layer Mamba blocks are first adopted and followed by a 1-D convolution layer to capture the local dependencies (Lee, 2016). The reason for utilizing the convolution layer is that the convolution operation can accommodate different realizations of the same underlying phone from various L2 speakers, thereby mitigating the temporal variability. The word-level representations $\mathbf{H}^{g_{L_w}^{wrd}}$ can be derived as follows:

$$\mathbf{H}'^{g_{L_w}^{wrd}} = \text{MambaBlock}_{wrd}(\mathbf{H}^{g_0^{wrd}}) \quad (12)$$

$$\mathbf{H}^{g_{L_w}^{wrd}} = \text{Conv1D}_{wrd}(\mathbf{H}'^{g_{L_w}^{wrd}}) \quad (13)$$

To obtain word-level aspect scores, we put $\mathbf{H}^{g_{L_w}^{wrd}}$ into the word-level APA module which contains three regressors to predict the word-level aspect scores $s_0^{g^{wrd}}, s_1^{g^{wrd}}, s_2^{g^{wrd}}$ (accuracy, stress, and total scores), respectively. To facilitate training efficiency, we propagate the word score to each of its phones during the training stage. In the inference phase, we ensure consistency by averaging the outputs corresponding to each word. In addition, $\mathbf{H}^{g_{L_w}^{wrd}}$ is viewed as $\mathbf{H}^{g_0^{utt}}$ for further modeling.

As for the utterance-level assessments, instead of prepending the `[CLS]` tokens to learn the utterance-level representation (Gong et al., 2022), we explore pooling-based approaches to aggregate the hidden information. To this end, we utilize an attention pooling layer similar to Peng et al. (2022). Specifically, assuming that a $d$-dimensional input sequence to the attention pooling layer is $\mathbf{h}_0, \mathbf{h}_1 \ldots, \mathbf{h}_{T-1}$, the pooling output is $\mathbf{h} = \sum_{i=0}^{T-1} \alpha_i \mathbf{h}_i$, where $\alpha_i$ is calculated by

$$\alpha_i = \frac{\exp\left(\mathbf{w}^T \mathbf{q}_i / \tau\right)}{\sum_{j=0}^{T-1} \exp\left(\mathbf{w}^T \mathbf{q}_j / \tau\right)} \quad (14)$$

where $\mathbf{w}$ is a learnable vector, $\mathbf{q}$ is the concatenated scores of $[s_0^{g^{phn}}, s_0^{g^{wrd}}, s_1^{g^{wrd}}, s_2^{g^{wrd}}]$, and $\tau$ is a controllable temperature hyperparameter. The whole process of utterance-level modeling can then be formulated by

$$\mathbf{H}^{g_{L_u}^{utt}} = \text{MambaBlock}_{utt}(\mathbf{H}^{g_0^{utt}}) \quad (15)$$

$$\mathbf{h}^{g^{utt}} = \text{AttentionPooling}_{utt}(\mathbf{H}^{g_{L_u}^{utt}}) \quad (16)$$

After obtaining $\mathbf{H}^{g_{L_u}^{utt}}$ from $L_u$-layer Mamba blocks, $\mathbf{h}^{g^{utt}}$ is derived through the attention pooling layer to predict the utterance-level aspect scores $s_0^{g^{utt}}, s_1^{g^{utt}}, s_2^{g^{utt}}, s_3^{g^{utt}}, s_4^{g^{utt}}$ (accuracy, completeness, fluency, prosody, and total scores) via an utterance-level APA module which contains five regressors corresponding to each score.

### 2.3 Optimization

**Automatic Pronunciation Assessment Loss:** In the proposed model, each APA module is optimized using Mean Square Error (MSE). The loss for multi-aspect multi-granular assessment, $\mathcal{L}_{APA}$, is calculated by assigning weights to each granularity level $g$:

$$\mathcal{L}_{APA} = \sum_{g \in G} \omega_g \cdot \frac{1}{N_g} \sum_{k=0}^{N_g-1} \cdot \mathcal{L}_{g_k} \quad (17)$$

where $\omega_g$ and $N_g$ are the tunable hyperparameter and number of aspect scores at granularity level $g$, respectively. $\mathcal{L}_{g_k}$ refers to the MSE loss computed for $k$-th aspect score at granularity level $g$.

**Mispronunciation Detection and Diagnosis Loss:** To be in line with previous MDD studies, our model incorporates canonical phoneme embeddings to enhance text prompt-awareness. Despite some performance improvements, the mismatch between the L2 learner's realized phones and canonical phones can still cause some deteriorating effects. This discrepancy can introduce inaccurate predictions that may potentially affect the overall quality of phonetic analysis. To mitigate this negative impact, we devise a new loss function tailored for the MDD task, as illustrated in Figure 3. Specifically, we first decouple the original cross-entropy loss into two separate losses, one for mispronunciations and the other for correct pronunciations:

$$\mathcal{L}_{Xent}^{mis} = -\sum_{t \in \mathcal{M}} \log(\hat{y}_t[y_t]) \quad (18)$$

$$\mathcal{L}_{Xent}^{hit} = -\sum_{t \in \mathcal{H}} \log(\hat{y}_t[y_t]) \quad (19)$$

where $\mathcal{M}$ and $\mathcal{H}$ are mispronunciation and correct pronunciation positions, respectively, and $\hat{y}_t[y_t]$ is the predicted probability of the true label $y_t$ at time step $t$. After obtaining two decoupled losses, the proposed decoupled cross-entropy loss (deXent) is obtained by the following formulation:

$$\mathcal{L}_{MDD} = \mathcal{L}_{Xent}^{hit} + \left(\frac{\mu^h}{\mu^m}\right)^\alpha \mathcal{L}_{Xent}^{mis} \quad (20)$$

where $\mu^m$ and $\mu^h$ denote the frequency of the mispronunciations and correct pronunciations in the training set, respectively, and $\alpha$ controls the weight magnitude. After that, we use $\mathcal{L}_{MDD}$ to optimize the MDD module, and the overall loss thus can be expressed by

$$\mathcal{L} = \mathcal{L}_{APA} + \beta \cdot \mathcal{L}_{MDD} \quad (21)$$

where $\beta$ is a tunable hyperparameter.

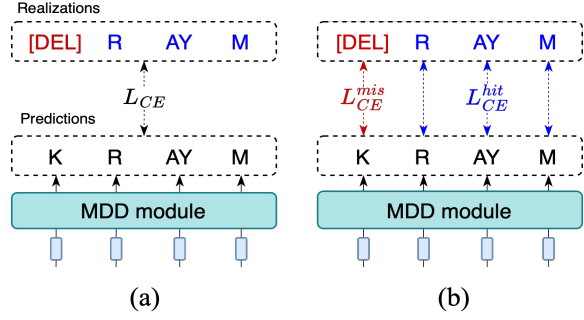

Figure 3: Difference between (a) the original cross-entropy loss and (b) the decoupled cross-entropy loss, given the text prompt "crime."

According to Equation 20, the proposed loss function can be viewed as one of the loss-balancing methods, such as focal loss (Lin et al., 2017) and class-balanced loss (Cui et al., 2019), to tackle the imbalance issue in MDD. However, in most end-to-end MDD methods, where the labels are phones instead of mispronunciations (0 or 1s), directly applying the existing loss-balancing methods on phones is implicit and can be sub-optimal when we aim to detect potential mispronunciations. Hence, we believe the proposed deXent provides a better alternative to end-to-end MDD.

| Model | Year | Phone Score | | Word Score (PCC) | | | Utterance Score (PCC) | | | | |
|---|---|---|---|---|---|---|---|---|---|---|---|
| | | MSE↓ | PCC↑ | Accuracy↑ | Stress↑ | Total↑ | Accuracy↑ | Completeness↑ | Fluency↑ | Prosody↑ | Total↑ |
| Deep Feature | 2021 | - | - | - | - | - | - | - | - | - | 0.720 |
| HuBERT Large | 2022 | - | - | - | - | - | - | - | 0.780 | 0.770 | - |
| Joint-CAPT-L1 | 2023 | - | - | - | - | - | 0.719 | - | 0.775 | 0.773 | 0.743 |
| LSTM | 2022 | 0.089 | 0.591 | 0.514 | 0.294 | 0.531 | 0.720 | 0.076 | 0.745 | 0.747 | 0.741 |
| GOPT | 2022 | 0.085 | 0.612 | 0.533 | 0.291 | 0.549 | 0.714 | 0.155 | 0.753 | 0.760 | 0.742 |
| 3M | 2022 | 0.078 | 0.656 | 0.598 | 0.289 | 0.617 | 0.760 | 0.325 | 0.828 | 0.827 | 0.796 |
| HiPAMA | 2023 | 0.084 | 0.616 | 0.575 | 0.320 | 0.591 | 0.730 | 0.276 | 0.749 | 0.751 | 0.754 |
| 3MH | 2023 | 0.071 | 0.693 | 0.682 | 0.361 | 0.694 | 0.782 | **0.374** | 0.843 | 0.836 | 0.811 |
| **HMamba** | 2024 | **0.062** | **0.739** | **0.708** | **0.366** | **0.718** | **0.807** | 0.278 | **0.848** | **0.843** | **0.829** |

Table 1: APA performance evaluations of our model and all strong baselines on the speechocean762 test set.

## 3 Experimental Setup

### 3.1 Dataset

We conducted experiments on speechocean762, a widely-used open-source dataset curated for APA and MDD research (Zhang et al., 2021). The dataset consists of 5,000 English-speaking recordings from 250 Mandarin L2 learners, divided evenly into training and test sets. For the APA task, pronunciation proficiency scores were assessed at various linguistic granularities and across different pronunciation aspects. Each score is evaluated by five experts using standardized rubrics. For the MDD task, the dataset provides an extra mispronunciation transcription annotated using a set of 46 phones. This set comprises 39 phones from the CMU dictionary[4], 6 L2-specific phones, and a [unk] token for unknown phones. Notably, there are no insertion errors in the utterances, and a [DEL] token is introduced to mark deletion errors of L2 learners. Therefore, the realized phones can be aligned with canonical phones in this dataset.

### 3.2 Training and Evaluation Details

**Training:** We optimized the model with Adam and a tri-phase rate scheduler (Baevski et al., 2020), where the learning rate was gradually increased during the first 40% of steps, held constant for the following 40%, and then linearly decayed for the remaining steps. The initial learning rate was set to 2e-3 except for the utterance-level APA module, which was set to 9e-5. Other implementation details are presented in Appendix A.

**Evaluation:** The evaluation metrics employed include the Pearson Correlation Coefficient (PCC) and Mean Square Error (MSE) for the APA task. On the other hand, we used precision, recall, F1-score, and phone error rate (PER) to evaluate the MDD performance, so as to be in accordance with prior studies. To ensure the validity of our experimental results, we conducted 5 independent trials for each experiment, running 20 epochs with different seeds. The metrics for each task are reported as the average of these trials.

### 3.3 Compared Baselines

For APA, we compare our proposed approach, HMamba, with various cutting-edge baselines which can be categorized into two families: single-aspect pronunciation assessment models or multi-granular multi-aspect pronunciation assessment models. The first group includes the Deep Feature (Lin et al., 2021b), HuBERT Large (Kim et al., 2022), and Joint-CAPT-L1 (Ryu et al., 2023). The second group encompasses LSTM, GOPT (Gong et al., 2022), 3M (Chao et al., 2022), HiPAMA (Do et al., 2023a), and 3MH (Chao et al., 2023). As for MDD, we compare HMamba with Joint-CAPT-L1, as to our knowledge it is the only attempt that jointly addresses the APA and MDD tasks with the speechocean762 dataset.

---
[4] http://www.speech.cs.cmu.edu/cgi-bin/cmudict

| Model | Phone Score | | Word Score (PCC) | | | Utterance Score (PCC) | | | | |
|---|---|---|---|---|---|---|---|---|---|---|
| | MSE↓ | PCC↑ | Accuracy↑ | Stress↑ | Total↑ | Accuracy↑ | Completeness↑ | Fluency↑ | Prosody↑ | Total↑ |
| LMamba | 0.071 | 0.694 | 0.678 | 0.299 | 0.689 | 0.790 | 0.234 | 0.844 | 0.838 | 0.816 |
| PMamba | 0.068 | 0.707 | 0.689 | 0.320 | 0.700 | 0.784 | 0.142 | 0.843 | 0.832 | 0.817 |
| **HMamba** | **0.062** | **0.739** | **0.708** | **0.366** | **0.718** | **0.807** | **0.278** | **0.848** | **0.843** | **0.829** |

Table 2: Performance comparison between different modeling structures.

| Model | Mispronunciations | | | PER ↓ |
|---|---|---|---|---|
| | Precision ↑ | Recall ↑ | F1 ↑ | |
| Joint-CAPT-L1 | 26.70% | **91.40%** | 41.50% | 9.93% |
| **HMamba** | **64.35%** | 63.41% | **63.85%** | **2.72%** |

Table 3: MDD performance evaluations of our model, compared with a representative multi-task approach (Ryu et al., 2023) on the speechocean762 test set.

| Loss | $\alpha$ | Mispronunciations | | | PER ↓ |
|---|---|---|---|---|---|
| | | Precision ↑ | Recall ↑ | F1 ↑ | |
| Xent | - | **77.07%** | 38.60% | 51.40% | **2.53%** |
| deXent | 0.3 | 70.06% | 54.10% | 61.04% | 2.61% |
| | 0.5 | 67.12% | 58.71% | 62.62% | 2.70% |
| | 0.7 | 64.35% | 63.41% | **63.85%** | 2.72% |
| | 0.9 | 57.74% | **71.12%** | 63.73% | 3.14% |

Table 4: Comparison of MDD performance between the original cross-entropy loss (Xent) and proposed decoupled cross-entropy loss (deXent).

## 4 Experimental Results and Discussion

### 4.1 APA Performance

**Overall Results:** In Table 1, we compare the APA performance of HMamba with other competitive baselines, leading to several key observations. Firstly, it is notable that our approach, HMamba, consistently outperforms all other methods on nearly all assessment tasks, particularly in terms of accuracy scores at phone, word, and utterance levels. This improvement stems from the joint modeling paradigm of APA and MDD, highlighting that pronunciation assessments can also benefit from phonetic error discovery, consistent with prior research findings (Ryu et al., 2023). In addition, by adopting SSL features, HMamba as well as other approaches like HuBERT Large, 3M, and 3MH, achieves significant improvements over the other APA methods in terms of utterance-level assessments. In comparison to other hierarchical models such as HiPAMA and 3MH, HMamba leverages an SSM structure instead of the Transformer structure, demonstrating superior performance on a variety of assessment tasks (further analysis between Mamba and Transformer are shown in Appendix C). In assessing utterance completeness, while HMamba falls behind 3M and 3MH, it is on par with HiPAMA. According to Zhang et al. (2021), the completeness refers to the percentage of the words that are actually pronounced. This may imply that our approach has a weaker ability to detect incompletely articulated words. However, this limitation could stem from the fact that HMamba focuses more on phoneme accuracy and mispronunciation detection, rather than purely evaluating word-level completeness.

**Effects of Hierarchical Structure:** To inspect the hierarchical structure that influences on the APA performance of the proposed approach, we conducted an experiment to replace the hierarchical structure with two other variants, resulting in two different models: LMamba and PMamba, respectively. LMamba has a similar structure to HMamba but outputs all assessment scores in the last layers regardless of their granular differences. On the other hand, PMamba adopts the parallel structure suggested by previous studies (Gong et al., 2022; Chao et al., 2022) that use prepended `[CLS]` tokens to predict utterance-level scores. According to the results shown in Table 2, HMamba outperforms PMamba and LMamba across all assessment aspects, highlighting the advantages of its hierarchical structure for the APA task. This finding aligns with previous research (Chao et al., 2023). In addition, the significant performance gaps between HMamba and LMamba also suggest that phone-level and word-level scores should be predicted in lower layers.

## 4.2 MDD Performance

**Overall Results:** We evaluate the MDD performance of HMamba by comparing it with another celebrated multi-task learning approach, Joint-CAPT-L1. As shown in Table 3, despite of lower recall rate, HMamba achieves a significant improvement in terms of F1-score over Joint-CAPT-L1, with an increase of 22.35%. Additionally, there is a marked reduction in PER by 7.21%. These substantial enhancements demonstrate that HMamba not only delivers accurate pronunciation assessments but also produces more robust and reliable mispronunciation detection and diagnosis results.

**Effects of Decoupled Cross-entropy Loss:** On the grounds of the distinct improvements in the MDD performance, we further analyze the underlying effects of our proposed decoupled cross-entropy loss (deXent). As illustrated in Table 4, training a text prompt-aware MDD model using the original cross-entropy loss (Xent) often yields high precision but low recall. This is because the model primarily relies on input canonical phones, leading it to predict prior phones and overlook the actual mispronunciations of a learner. Such a model may not be suitable for educational settings where accurately detecting potential mispronunciations is critical. To remedy this, the proposed deXent method sufficiently provides a feasible solution. By adjusting the weighting factor $\alpha$, we can better strike the balance between precision and recall, thus optimizing the F1-score. This flexibility is particularly vital in CAPT, where both over-detection and under-detection of mispronunciations can severely disrupt the learning process—a challenge often neglected by most existing end-to-end methods. While adopting deXent may result in a minor increase in PER, this slight performance tradeoff is justifiable for the significant gains in overall MDD effectiveness.

**Limitations of Decoupled Cross-entropy Loss:** In Table 3, the MDD performance of HMamba is reported based on maximizing the F1 score using the deXent, as we believe both the precision and recall metrics are critical for MDD. However, a potential limitation of using deXent as the loss function is that, while it may help balance precision and recall for MDD, it may not simultaneously improve both metrics. This limitation likely stems from the close relationship between the mispronunciation distribution and the loss-balancing mechanism of deXent.

## 5 Conclusion

In this paper, we have presented a novel hierarchical selective state space model (dubbed HMamba) for multifaceted CAPT application. Extensive experimental results substantiate the viability and efficacy of the proposed method compared to several top-of-the-line approaches in terms of both the APA and MDD performance. In future work, we envisage mitigating the issue of data imbalance from an optimization perspective. In addition, another key area for future research involves tackling the assessment of open-response scenarios in CAPT.

## Limitations

**Lack of Accent Diversity.** The dataset used in this study comprises only Mandarin L2 learners, which would probably limit the generalizability of the proposed model. As a result, it might be inapplicable when assessing L2 learners with diverse accents. This lack of accent diversity could lead to biases and inaccuracies in pronunciation assessment for learners from different linguistic backgrounds.

**Limited Interpretability.** The proposed model is designed to replicate expert annotations without relying on manual assessment rubrics or external knowledge databases, which would make it challenging to provide clear and reasonable explanations for the assessment results. This insufficiency of interpretability might hinder its acceptance and trustworthiness among educators and learners who require transparent and justifiable assessments.

**Limited Generalizability** This research is centered on the "reading-aloud" pronunciation training scenario, where it is assumed that the L2 learner accurately pronounces a predetermined text prompt. This would narrow the applicability of our models to other learning contexts, such as spontaneous speech or open-ended conversations.

## Ethics Statement

We acknowledge that all co-authors of this work comply with the ACL Code of Ethics and adhere to the code of conduct. Our experimental corpus, speechocean762, is widely used and publicly available, and we believe there are no potential risks associated with this work.

## Acknowledgments

This work was supported by the Language Training and Testing Center (LTTC), Taiwan. Any findings and implications in the paper do not necessarily reflect those of the sponsor.

|  | APA | | | | | | | | | MDD | |
|---|---|---|---|---|---|---|---|---|---|---|---|
|  | Phone | | Word | | | Utterance | | | | | |
|  | MSE↓ | Acc.↑ | Acc.↑ | Stress↑ | Total↑ | Acc.↑ | Comp.↑ | Fluency↑ | Prosody↑ | Total↑ | F1↑ | PER↓ |
| **HMamba** | **0.062** | **0.739** | **0.708** | **0.366** | **0.718** | **0.807** | 0.278 | **0.848** | **0.843** | **0.829** | **63.85%** | **2.72%** |
| Acoustic Features | | | | | | | | | | | | |
| -wav2vec2 | 0.062 | 0.736 | 0.708 | 0.326 | 0.718 | 0.801 | 0.185 | 0.840 | 0.833 | 0.823 | 63.63% | 2.79% |
| -HuBERT | 0.063 | 0.735 | 0.706 | 0.344 | 0.715 | 0.804 | 0.216 | 0.843 | 0.838 | 0.825 | 63.49% | 2.90% |
| -wavLM | 0.063 | 0.731 | 0.705 | 0.355 | 0.715 | 0.806 | 0.247 | 0.844 | 0.838 | 0.827 | 63.39% | 2.97% |
| -duration | 0.063 | 0.734 | 0.705 | 0.341 | 0.715 | 0.804 | **0.299** | 0.844 | 0.838 | 0.826 | 63.70% | 2.80% |
| -energy | 0.063 | 0.735 | 0.706 | 0.358 | 0.716 | 0.802 | 0.257 | 0.840 | 0.834 | 0.823 | 63.28% | 2.78% |
| -GOP | 0.066 | 0.719 | 0.699 | 0.293 | 0.706 | 0.795 | 0.228 | 0.837 | 0.829 | 0.817 | 61.72% | 2.79% |
| Phonological Features | | | | | | | | | | | | |
| -absolute-pos | 0.063 | 0.735 | 0.706 | 0.332 | 0.715 | 0.802 | 0.261 | 0.843 | 0.838 | 0.825 | 63.48% | 2.79% |
| -relative-pos | 0.063 | 0.733 | 0.704 | 0.352 | 0.714 | 0.804 | 0.220 | 0.847 | 0.841 | 0.825 | 62.74% | 2.81% |
| -canonical-phn | 0.083 | 0.624 | 0.604 | 0.310 | 0.617 | 0.775 | 0.147 | 0.842 | 0.836 | 0.801 | 28.06% | 14.55% |

Table 5: Feature ablations of HMamba (MDD performance is reported with F1 and PER as representative metrics).

# Appendix

## A Implementation Details

**Feature Extraction:** We adopt an open-source acoustic model[5] to extract GOP features, which also serves as an aligner for force alignment. Subsequently, the phone-level duration, energy statistics, and SSL features (average over time frames) are computed according to the alignment. The resulting acoustic features $X$ and all embeddings are 128 dimensions. For all Mamba blocks, we set the number of hidden units to 128 and use a kernel size of 4 for the 1-D convolution. The SSM modules follow the original configuration used in Mamba. $L_p, L_w, L_u$ are set to 3, 1, 1, respectively. In addition, the word-level 1-D convolution has 256 kernels, each with a size of 3.

**Hyperparameters setting:** $\tau$ in attention pooling layer is set to 1.0. The combining weights $\omega_g$ for APA loss are uniformly set to 1.0 for each granularity level $g$. Parameters $\alpha$ and $\beta$ are tuned to be 0.7 and 0.003, respectively.

---
[5] https://kaldi-asr.org/models/m13

## B Feature Ablations

In Table 5, we conduct an ablation study on the feature extraction to inspect the factors that influence on APA. Specifically, we removed one factor at a time to investigate the performance variations.

**Acoustic Features:** According to the ablation experiment, each acoustic feature used in this work contributes to specific aspect assessments. While the model may perform better in assessing utterance completeness without phone duration features, the other aspect assessments and MDD performance decreases synchronously. Furthermore, among all acoustic features, GOP is the most crucial factor in relation to both of the APA and MDD performance.

**Phonological Features:** As for the phonological features, the canonical phoneme embeddings are the most critical features overall, particularly in MDD. Without canonical phoneme embeddings, the performance dramatically degrades in F1 score and PER. Since the number of mispronunciations is typically far less than the correct pronunciations, canonical phoneme embeddings can provide ample text-prompt information to complement the acoustic cues. The impact of canonical phoneme embeddings on APA is consistent with Gong et al., 2022, and we take it a step further in this work by demonstrating that they are also pivotal for MDD.

| Block Type | APA | | | | | | | | | | MDD | |
|---|---|---|---|---|---|---|---|---|---|---|---|---|
| | Phone | | Word | | | Utterance | | | | | F1↑ | PER↓ |
| | MSE↓ | Acc.↑ | Acc.↑ | Stress↑ | Total↑ | Acc.↑ | Comp.↑ | Fluency↑ | Prosody↑ | Total↑ | | |
| Transformer | 0.071 | 0.692 | 0.689 | 0.294 | 0.700 | 0.797 | 0.165 | 0.844 | 0.839 | 0.819 | 60.14% | 3.50% |
| **Mamba** | **0.062** | **0.739** | **0.708** | **0.366** | **0.718** | **0.807** | **0.278** | **0.848** | **0.843** | **0.829** | **63.85%** | **2.72%** |

Table 6: Performance comparison between Mamba block and Transformer block.

| Block Type | Params(M)↓ | MACs(G)↓ |
|---|---|---|
| Transformer | 1.469 | 3.806 |
| **Mamba** | **1.141** | **2.954** |

Table 7: Computational efficiency between Mamba block and Transformer block.

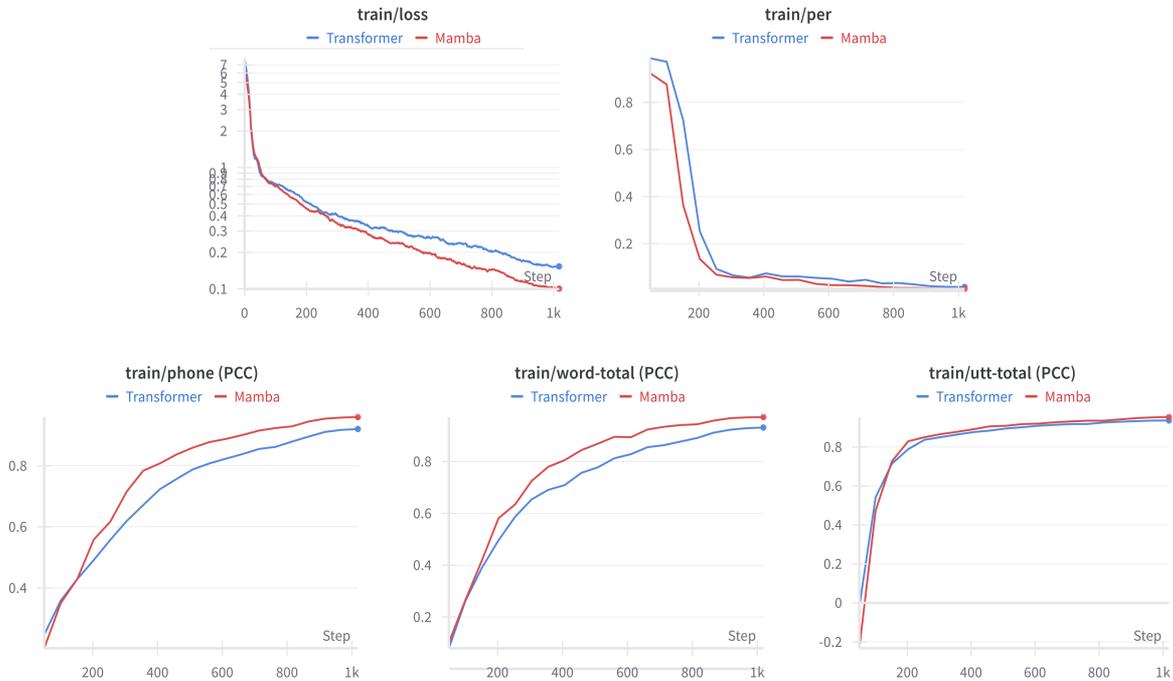

Figure 4: Comparison of the training curves for models equipped with Mamba and Transformer.

## C Mamba v.s. Transformer

To validate the effectiveness of Mamba over Transformer in different facets, we replace each Mamba block in HMamba with a vanilla Transformer block (encoder only). In the following, we perform a set of qualitative analyses for comparisons of these two variant structures.

**Performance comparison:** In the first experiment, we compare the APA and MDD performance by utilizing Mamba or Transformer as a basic block, respectively. According to the results shown in Table 6, Mamba consistently outperforms Transformer across all assessment tasks and MDD, especially in phone- and word-level assessments. The key difference between Mamba and Transformer is that Mamba leverages 1-D convolution to model the local context dependency which has been shown crucial in either APA (Do et al., 2023a) or MDD (Lee, 2016) task. These findings also align with other research in the speech community, such as speech separation (Jiang et al., 2024) and speech enhancement (Zhang et al., 2024).

**Computational efficiency:** We further investigate the computational efficiency of two variants of architectures with the number of their parameters and multiply-accumulate operations (MACs). In Table 7, we observe that the model equipped with Mamba has fewer parameters and requires fewer MACs compared to the model with Transformer. This reduced complexity suggests that Mamba is not only more resource-efficient but also potentially more scalable for practical applications where both performance and resource constraints are critical.

**Training efficiency:** To track the full training process of the two architectures, we plot various training curves in Figure 4, including the training loss (log scale) and the changes in key metrics, such as PER and the total scores of the phones, words, and utterances. Mamba (red) consistently exhibits lower training loss than Transformer (blue) throughout the steps, suggesting that Mamba enables faster and better convergence, ultimately achieving a smaller loss overall. As training progresses, Mamba rapidly outperforms Transformer in terms of PER and PCC for both phone and word total scores. Although this advantage is less pronounced in the PCC curve for the utterance total score, Mamba still surpasses Transformer to a moderate extent.